\documentstyle[12pt]{article}
\begin{document}
\baselineskip 10pt plus .1pt minus .1pt
\pagestyle{empty}
\begin{center}
{\large\bf{Degenerate Dirac Neutrinos In An ${SU(2)}_L\times
{U(1)}_Y$ Model With $S_3\times Z_3\times Z_4$ Symmetry}}
\end{center}
\vskip 1in
\begin{center}
{\bf Ambar Ghosal and Asim K. Ray}
\vskip .25in
{Department of Physics\\
Visva-Bharati University\\
Santiniketan 731 235\\
India}
\end{center}
\vskip .5in
\begin{center}
{\bf Abstract}
\end{center}
\vskip .5in
We demonstrate that almost degenerate Dirac neutrinos of  mass
of the order of a few eV and transition magnetic  moment  of  the
order of ${10}^{-11}$ $\mu_B$  can be obtained in an
${SU(2)}_L\times {U(1)}_Y$ model with $S_3\times Z_3\times Z_4$
discrete symmetry and appropriate Higgs. Transition magnetic moment
of the Dirac neutrino arises
from the contribution of leptons and  charged  Higgs
fields at the one loop level.
\newpage
\baselineskip 21pt plus .1pt minus .1pt
\pagestyle{plain}
\setcounter{page}{2}
\noindent
Three possible scenarios of neutrino masses have been proposed by Caldwell and
Mohapatra [1] which can simultaneously reconcile the solar [2]
and atmospheric [3] neutrino deficits on earth as well as the apparent
need for neutrino as a hot dark matter component [4]. Two of
them contain a sterile neutrino besides the three
neutrinos $\nu_e, \nu_\mu$ and $\nu_\tau$.
The scenario, which does not contain any
sterile neutrino, requires the three neutrinos
$\nu_e, \nu_\mu$ and $\nu_\tau$ to be almost degenerate in mass of the order
of 2-3 eV. Furthermore, the recent results from the chromium (${}^{51}$Cr)
source experiment [5] carried out by the GALLEX collaboration
implies interesting limits on the parameters $\Delta m^2$ and
$sin^2 2\theta$ describing the neutrino flavour oscillation. Values
of $\Delta m^2 > 0.17 eV^2$ for maximal mixing and of $sin^2
2\theta > 0.38$ for $\Delta m^2 > 1\, eV^2$ are ruled out (at 90
$\%$c.l.) [6]. Another result from Bugey [7] sets the minimum
excluded values of $\Delta m^2$ and $sin^2 2\theta$ are $1\times
{10}^{-2} eV^2$ and $2\times {10}^{-2}$ (at 90 $\%$ c.l.)
respectively. Although the recent experimental results have
strengthened the conjecture of the neutrino flavour oscillation,
but the phenomenon is yet to be established [8].\\

Assuming the neutrino flavour oscillation,
the hierarchy of the
masses of the three neutrinos is proposed on the
phenomenological basis [9, 10] as
${m_{\nu_1}^2 ={m_o}^2}$, ${m_{\nu_2}^2 ={m_o}^2}+ \Delta_{21}$,
and ${m_{\nu_3}^2 ={m_o}^2} + \Delta_{32} + \Delta_{21}$
with ${m_0}^2 >> \Delta_{21}, \Delta_{32}$ where
$\Delta_{ij} ={m_{\nu_i}}^2 - {m_{\nu_j}}^2$.
Moreover, if $\nu_e \rightarrow \nu_\mu$ and
$\nu_\mu \rightarrow \nu_\tau$
oscillations are
assumed to be responsible for the solar and atmospheric  neutrino
deficits respectively, then $|\Delta_{32}|>>|\Delta_{21}|$.
The degeneracy between the three generations is lifted due the presence of
$\Delta_{ij}$ terms which are expected to be small. The small Majorana neutrino
mass of the order of few eV can be realized through the introduction of
intermediate or large (GUT scale) mass scale [11] emerging from the
symmetry breaking of the model through the standard see-saw mechanism [12].
However, the Dirac neutrinos are expected to be heavy (as the masses of the
Dirac neutrinos are proportional to the electroweak symmetry breaking scale)
unless the associated Yukawa couplings are exteremely small.\\

Another interesting aspect of neutrino physics is to accommodate large
magnetic moment of the order of ${10}^{-11} \mu_B$ as suggested
by Okun, Voloshin and Vysotsky [13] in order to explain the
anticorrelation between the Sun
spot activity and the observed solar neutrino flux on earth.
The bound obtained for the neutrino magnetic moment from the laboratory
experiments of $\bar{\nu_e} e$
scattering $(< 1.5\times {10}^{-10}\mu_B)$ [14], which has been
corrected to $(<4\times {10}^{-10}\mu_B)$ [15], is just below the above
mentioned value.
{}From the
astrophysical    consideration     of     steller     cooling,
a much more stronger limit is derived
$(\mu_{\nu_e}<(0.3-1.0)\times {10}^{-11}\mu_B)$ [16]. In order to explain
the solar neutrino deficits, two types of oscillations are
suggested [17, 18].
The first one $\nu_{eL}\rightarrow \nu_{eR}$,  is
supported due to the apparent anticorrelation between the Sun
spot activity and the observed neutrino flux on earth, however, this is in
conflict with the observation of SN1987 A data for
the energy loss of the supernova burst [19]. The other
scenario, which assumes the oscillation of
$\nu_{eL}\rightarrow {(\nu_\mu^R)}^c$
seems to be a much more reasonable choice, since it has no
severe bounds from the observation of SN1987 A as well as nucleosynthesis.\\

In the present work  we  demonstrate that, within the framework of
Standard $SU(2)_L\times U(1)_Y$ model with $S_3\times Z_3\times Z_4$
discrete symmetry, right-handed neutrinos and appropriate Higgs
fields, almost degenerate Dirac neutrinos
and a large transition magnetic moment
${10}^{-11}\mu_B$
can be achieved through the incorporation of the two widely different lepton
number symmetry breaking scales.
Our present model admits light degenerate Dirac neutrinos of mass of the
order of 1 eV, contrary to the expectation of heavy Dirac neutrinos,
through the effective coupling of the Higgs fields.
The first and second generations are completely degenerate in mass and
the third generation can also be made almost degenerate with the
others through some appropriate fine tuning
of the model parameter. The large transition magnetic moment
$(\nu_{e_L}\rightarrow \nu_{\mu_R})$, consistent with the
phenomenological bounds from SN1987 A and nucleosynthesis as metioned earlier,
arises due to the
quartic couplings of the Higgs fields through the charged Higgs exchange
at the one loop level.\\

We concentrate on the leptons and Higgs fields of the model. The lepton
content in the present model is as usual\\
\noindent
$$l_{iL}(2, -1, 1) = {\pmatrix{\nu_e\cr
                               e}}_{iL},
\, \nu_{iR}(1, 0, 1)\eqno(1)$$
\noindent
where i=1,2,3 is the generation index. The following Higgs
fields are considered with the Vacuum Expectation Values (VEV) as indicated\\
$$\phi_i(2, 1, 0) = \pmatrix{0\cr
                             v_i},
\, \eta_j(1, 0, x) = k_j\eqno(2)$$
\noindent
where $i = 1,...5$ and $j = 1,..4$ are the number of
the Higgs doublets and singlets respectively. The last digit in the parenthesis
represents lepton number $L(= L_e + L_\mu + L_\tau)$. The lepton number of the
singlet Higgs fields is arbitrary non-zero number since, the singlets do
not couple with the leptons at the tree level due to the discrete
$Z_3\times Z_4$
symmetry incorporated in the model. In this sense, our present model is
different from the well known CMP majoron model [20]. However, the present
model also admits spontaneous lepton number violation through the development
of non-zero VEV's of the singlet Higgs fields like the CMP majoron model.\\

The leptons and the Higgs fields transform under discrete
$S_3\times Z_3\times Z_4$ symmetry as follows:\\
\noindent
i) $S_3$ symmetry :\\
$$(l_{1L}, l_{2L})\rightarrow 2, l_{3L}\rightarrow 1,
(\nu_{\mu R}, \nu_{eR})\rightarrow 2, \nu_{\tau
R}\rightarrow 1,$$
$$(e_R , \mu_R)\rightarrow 2, \tau_R\rightarrow 1,
\phi_1 \rightarrow 1, \phi_2 \rightarrow 1, \phi_3 \rightarrow 1,$$
$$(\phi_4, \phi_5)\rightarrow 2,
\eta_1 \rightarrow 1, \eta_2 \rightarrow 1, \eta_3 \rightarrow 1,
\eta_4 \rightarrow 1\eqno(3)$$
\noindent
ii) $Z_3\times Z_4$ symmetry:\\
$$(l_{1L}, l_{2L})\rightarrow (l_{1L}, l_{2L}), l_{3L}\rightarrow l_{3L},
(\nu_{\mu R}, \nu_{eR})\rightarrow (\nu_{\mu R}, \nu_{eR})$$
$$\nu_{\tau R}\rightarrow \nu_{\tau R},
(e_R , \mu_R)\rightarrow i\omega(e_R , \mu_R),
\tau_R\rightarrow i\omega \tau_R,
\phi_1\rightarrow \phi_1,$$
$$\phi_2\rightarrow \omega\phi_2, \phi_3\rightarrow -i\omega^\star\phi_3,
(\phi_4, \phi_5)\rightarrow -i\omega^\star(\phi_4, \phi_5),$$
$$\eta_1\rightarrow i\omega\eta_1,
\eta_2\rightarrow i\omega^2\eta_2,
\eta_3\rightarrow -\omega\eta_3,
\eta_4\rightarrow -i\omega^\star\eta_4\eqno(4)$$
\noindent
where $\omega = e^{{2\pi i\over 3}}$.\\

The present model admits only the Dirac neutrino mass terms. The discrete
symmetry prohibits the Majorana mass terms in the lepton-Higgs Yukawa
interactions as well as gives rise to some vanishing elements in the
Dirac neutrino mass matrix (at the tree level).
 We also discard any explicit lepton number
violating terms in the present model.
The   purpose   of
incorporation of $S_3$ permutation symmetry is to generate the equality
between the Yukawa couplings and VEV's of the Higgs fields in order to
get degenerate neutrino masses.
Several other interesting applications of $S_3$ symmetry have been investigated
in the context of quark mass matrices and CP phenomenology [21]. The Higgs
field $\phi_1$ couples with the neutrinos and is prohibited to couple with
the charged leptons. The $\phi_2$ Higgs field, although neither
couples with the neutrinos
nor with the charged leptons directly, but, gives rise to small degenerate
neutrino masses through its coupling with the $\phi_1$, $\eta_1$ and
$\eta_2$ Higgs
fields. The purpose of incorporation of $\phi_3$ Higgs
field, which couples with the charged leptons, is to generate
large transition magnetic moment through the coupling with
 $\phi_1$, $\eta_3$ and $\eta_4$
Higgs fields similar to the diagram given in Ref.18. The doublet $(\phi_4,
\phi_5)$ is necessary
to achieve non-degenerate charged lepton mass matrix. It is to be noted that,
without the doublet $(\phi_4, \phi_5)$, the charged lepton mass matrix will
also becomes degenerate, which is unphysical. The four Higgs
singlets are necessary to achieve the lepton number invariant terms in the
Higgs potential and we will
show that they play an important role to generate small Dirac neutrino
masses and large transition magnetic moment.\\

Another interesting feature in our present model is that the
absence of coupling at the tree level between the majorons
(generating due to the development of non-zero VEV's of the singlets through
spontaneous lepton number violation) and the right-handed
neutrinos forbids the neutrino decays
$(\nu_h\rightarrow \nu_l J)$ and/or annihilation of neutrinos
$(\nu\nu\rightarrow JJ)$ which leads to a longer life time of the neutrinos
so that the
neutrinos could be a relevant part of the dark matter [22].\\

The most general renormalizable Higgs potential, respecting
$S_3\times Z_3\times Z_4$ discrete symmetry, in the present model can be
written as\\
$$V = V(\phi_i) + V(\eta_j) + V(\phi_i, \eta_j)\eqno(5)$$
Explicitly the terms are given as follws :\\
$$V(\phi_i) = -[\sum_{i=1}^{3} m_i^2(\phi_i^\dagger\phi_i) +
  [\sum_{i=1}^{3} \mu_i{(\phi_i^\dagger\phi_i)}^2
-m_4^2(\phi_4^\dagger\phi_4 +\phi_5^\dagger\phi_5) +$$
$$\mu_4[{(\phi_4^\dagger\phi_4)}^2 +{(\phi_5^\dagger\phi_5)}^2] +
\lambda_1(\phi_1^\dagger\phi_1)(\phi_2^\dagger\phi_2) +
\lambda_2(\phi_1^\dagger\phi_1)(\phi_3^\dagger\phi_3) +$$
$$\lambda_3(\phi_1^\dagger\phi_1)(\phi_4^\dagger\phi_4 +\phi_5^\dagger\phi_5)
+ \lambda_4(\phi_2^\dagger\phi_2)(\phi_3^\dagger\phi_3) +
\lambda_5(\phi_2^\dagger\phi_2)(\phi_4^\dagger\phi_4
+\phi_5^\dagger\phi_5)$$
$$+\lambda_6(\phi_3^\dagger\phi_3)(\phi_4^\dagger\phi_4
+\phi_5^\dagger\phi_5)$$
$$+\lambda_7(\phi_4^\dagger\phi_5\phi_4^\dagger\phi_3 +
\phi_3^\dagger\phi_4\phi_5^\dagger\phi_4 +
\phi_5^\dagger\phi_4\phi_5^\dagger\phi_3 +
\phi_3^\dagger\phi_5\phi_4^\dagger\phi_5)\eqno(6)$$

$$V(\phi, \eta) = [\sum_{i=1,..5}^{j=1,..4}\lambda_{ij}(\phi_i^\dagger\phi_i)
(\eta_j^\star\eta_j)\, (with\,  \lambda_{4j}=\lambda_{5j})
+\lambda^\prime(\phi_1^\dagger\phi_2\eta_2^\star\eta_1 +
\eta_1^\star\eta_2\phi_2^\dagger\phi_1)$$
$$+\lambda (\phi_1^\dagger\phi_3\eta_3^\star\eta_4 +
\eta_4^\star\eta_3\phi_3^\dagger\phi_1) +
\lambda^{\prime\prime}(\phi_2^\dagger\phi_3\eta_2^\star\eta_3 +
\eta_3^\star\eta_2\phi_3^\dagger\phi_2)\eqno(7)$$
\noindent
where we have neglected $V(\eta)$ part of the potential, since, it is
not necessary for our present analysis. Substituting the VEV's of the Higgs
fields in Eqn.(6) and (7) and minimizing with respect to $v_1$, we get
$$v_1 = -{B\over A}\eqno(8)$$
\noindent
with $$B= (\lambda^\prime v_2 k_2 k_1 + \lambda v_3 k_3 k_4)$$
$$A = - m_1^2 + \lambda_1 v_2^2 + \lambda_2 v_3^2 + \lambda_3(v_4^2 + v_5^2)
+ \lambda_{11}k_1^2 + \lambda_{12}k_2^2 +\lambda_{13}k_3^2 + \lambda_{14}
k_4^2$$
\noindent
where we have neglected $\mu_1$ term for simplicity.\\

On simplification of Eqn.(8), we obtain \\
$$v_1 = {{\lambda^\prime v_2 k_2 k_1 + \lambda v_3 k_3 k_4}\over
m_1^2}\eqno(9)$$
\noindent
assuming $m_1^2$ is much larger than all other $\lambda$'s appeared in the
denominator of Eqn.(8). It is to be noted that the above
assumptions do not lead to any drastic changes in the result of
our present analysis. Similarly, minimizing the Higgs potential with respect
to $v_2$ and $v_3$, we get\\
$$v_2 = {{\lambda^{\prime\prime} v_3 k_2 k_3 + \lambda^\prime v_1 k_2 k_1}\over
m_2^2}\eqno(10)$$
and\\
$$v_3 = {{\lambda v_1 k_3 k_4 + \lambda^{\prime\prime} v_2 k_2 k_3}\over
m_3^2}\eqno(11)$$
\noindent
We further consider that $v_3$ = 0, and, thus, from Eqn.(9) we get\\
$$v_1= {{\lambda^\prime v_2 k_1 k_2}\over m_1^2}\eqno(12)$$
\noindent
It is to be noted that the quartic coupling $\lambda^\prime$, which generates
large magnetic moment, will also contribute to the neutrino masses unless we
choose $v_3$=0. However, our choice is not unnatural [18],infact, by an
orthogonal
transformation between the $\phi_3$, $\phi_4$ and $\phi_5$ Higgs fields,
the vacuum structure can be
arranged consistently with the choice $v_3 = 0$.
Furthermore, we infer the hierarchy of the VEV's of the
Higgs fields as \\
$$k_3 , k_4\gg v_i > k_1 , k_2\eqno(13)$$
\noindent
because of the fact that the VEV's of the singlets associated with $\lambda$
term should be much more larger than the VEV's of the singlets associated
with $\lambda^\prime$ term to yield large magnetic moment and small neutrino
masses. Thus, it is necessary to incorporate two widely different lepton number
symmetry
breaking scales, two of them $(\sim 1 TeV)$ are much above the electroweak
symmetry breaking scale and the other two scales $(\sim 100 GeV)$ are
near to the electroweak scale. Moreover, all the lepton
number violating processes at low energy (such as $\mu\rightarrow eee$,
$\mu\rightarrow e\gamma$, $K_L\rightarrow \mu e$ etc.)
are highly suppressed due to the small mass squared differences of neutrinos.
Such type of scenario has been investigated in
Ref.23 in the context of Baryogenesis.\\

The most general $S_3\times Z_3\times Z_4$ discrete symmetry
invariant lepton-Higgs Yukawa interaction, in our present model is as follows\\
$$-L_Y = [f_1(\bar{l_{1L}}\nu_{\mu R} + \bar{l_{2L}}\nu_{eR})
+ f_2 \bar{l_{3L}}\nu_{\tau R}]\tilde{\phi_1} +
g_1(\bar{l_{1L}} e_R + \bar{l_{2L}}\mu_R)\phi_3 +$$
$$g_2(\bar{l_{1L}} \tau_R \phi_4+ \bar{l_{2L}}\tau_R\phi_5)+
g_3 \bar{l_{3L}} \tau_R \phi_3 +
g_4(\bar{l_{3L}} e_R\phi_4 + \bar{l_{3L}}\mu_R\phi_5)+$$
$$g_5(\bar{l_{1L}} \mu_R\phi_4 + \bar{l_{2L}}e_R\phi_5) + h.c.\ .\eqno(14)$$
\noindent
Substituting the VEV's of the Higgs fields in Eqn.(14), the Dirac neutrino mass
matrix turns to be \\
$$M_\nu^D = \pmatrix{0&a&0\cr
                     a&0&0\cr
                     0&0&\xi a}\eqno(15)$$

where $a=f_1 v_1$, $\xi = {f_2\over f_1}$.\\
\noindent
It is to be mentioned that the mixing between the third and
the other two generations
has made zero (at the tree level ) by choice and this is in agreement with the
ansatz for $M_\nu^D$
in Ref.1. The non-zero contribution to the vanishing elements
of $M_\nu^D$ arise at the one loop level and gives rise to a
very small mixing angles consistent with the present
experimental limits as mentioned earlier.
Diagonalizing $M_\nu^D$, we obtain the mass eigenvalues of the three
Dirac neutrinos as\\
$$m_{\nu_1} = m_{\nu_2} = a = m_0\eqno(16)$$
$$m_{\nu_3} = \xi m_0\eqno(17)$$
\noindent
Thus we see that the present model admits two degenerate Dirac neutrinos and
the degeneracy between $m_{\nu_3}$ and $m_{\nu_1}$ or $m_{\nu_2}$ is lifted
due to the presence of the factor $\xi$. We will estimate the value of $\xi$
from the
knowledge of the experimental data of atmospheric neutrino, in the following.
Substituting Eqn.(12) in Eqn.(16), we obtain the mass of the two degenerate
Dirac
neutrinos as\\
$$m_{\nu_1} = m_{\nu_2} = {{f_1\lambda^\prime v_2 k_1 k_2}\over
m_1^2}\eqno(18)$$

\noindent
For a generic choice of model parameters $v_2$ = 100 GeV, $k_1 =
k_2$ = 100 GeV,
$m_1 \sim $ 1 TeV and $f_1 \sim $ 1 we obtain $m_{\nu_1} = m_{\nu_2}$ = 1 eV
for $\lambda^\prime\sim {10}^{-9}$. Such a small coupling is also consistent
with the other area of investigations [24]. However, the model contains a tiny
parameter space and there is not much freedom in the variation of model
parameters.
In particular, $v_2$ is restricted in the range (100 - 250) GeV and,
although $k_1, k_2$ can be varied in a wide range but, to yield
small neutrino masses, $k_1, k_2 \geq $ 100 GeV [18, 22].
The Yukawa coupling $f_1$ and the coefficient of the Higgs potential
$\lambda^\prime$ are less than equal to unity in order to satisfy the unitarity
bound.
The mass of the neutral Higgs boson $m_1$ is restricted within the range
(65.1 GeV - 1 TeV) where the lower bound comes from the results
of four combined experiments at CERN [25] whereas the
upper bound is also due to the unitarity of the theory.\\

The $\xi$ factor, which lifts the degeneracy between the three neutrinos
,can be determined from the atmospheric neutrino problem.
The hierarchy between $m_0$ and $m_{\nu_3}$ is manifested from the value of
$\Delta_{32} = m_0^2(\xi^2 - 1)$. For a typical value of
$\Delta_{32}\sim 4 \times {10}^{-3}$ $eV^2$ [7] which can explain the
atmospheric
neutrino deficit, we get $\xi \sim 1$ and thus, from
Eqn. (17), we obtain $m_{\nu_3}\sim m_0$.\\

The present model admits a large transition magnetic moment due to the presence
of the
quartic coupling term $\lambda (\phi_1^\dagger\phi_3\eta_3^\star\eta_4 +
\eta_4^\star\eta_3\phi_3^\dagger\phi_1)$ in the Higgs potential through charged
Higgs exchange at the one loop level. Similar diagram
has also been obtained in Ref.18 in which the internal fermion lines
are not ordinary leptons, contrary to the present model. The contribution
to $\mu_\nu$ in the present model
is given by (in the weak basis)\\
$$\mu_\nu\sim {e\over 8\pi^2} {f_1 g_1} {{m_e k_3 k_4\lambda }\over
{{m_{\phi_1}^+}^2 {m_{\phi_3}^+}^2}}\eqno(19)$$
\noindent

where $m_{\phi_1}^+$ , $m_{\phi_3}^+$ are the masses of the charged Higgs
bosons.
With the
following choice of model parameters consistent with the present
experimental limits [25] $m_{\phi_1}^+\sim $ 200 GeV,
$m_{\phi_3}^+\sim$ 100 GeV, $k_3\sim k_4\sim$ 1 TeV, present experimental bound
on $\mu_\nu< {10}^{-11}\mu_B$ can be obtained for $f_1 g_1 \lambda\sim 1$.\\

It is to be noted that the $\lambda$ term, which contributes to the
neutrino magnetic
moment, cannot contribute to the neutrino masses
due to our choice of $v_3$ = 0
and, hence, the lepton number breaking scales,which are much higher
than the electroweak
scale, do not give rise to any contribution to the neutrino masses.
Similarly, the $\lambda^\prime$ term of the Higgs potential which generates
small neutrino masses is prohibited to contribute to the magnetic moment due
to the discrete symmetry incorporated in the model. Thus, by
decoupling the two sets of widely different lepton number
symmetry breaking scales, it is possible to achieve small degenerate Dirac
neutrinos $\sim$ 1 eV and large transition magnetic moment $\sim
{10}^{-11}\mu_B$
in the present model. Furthermore, $\nu_{eL}\rightarrow \nu_{\mu R}$
flipping in the magnetic field of the Sun requires the mass
splitting
$\Delta_{21}\leq {10}^{-4} {eV}^2$ [26]. In the present model
$\Delta_{21}$ vanishes due to the degeneracy in mass between
$\nu_e$ and $\nu_\mu$. However, a small mixing (which in turn
generates non-zero $\Delta_{21}$) can be obtained from the one
loop level and is expected to be well within the present
experimental limit.\\

In summary, we demonstrate that Standard $SU(2)_L\times U(1)_Y$ model
with $S_3\times Z_3\times Z_4$ discrete symmetry, right-handed neutrinos and
appropriate Higgs fields can give rise to small degenerate Dirac neutrinos
$\sim$ 1 eV which can simultaneously reconcile the solar and atmospheric
neutrino deficits as well as the candidature of the neutrino as a hot dark
matter,
through spontaneous lepton number violation at two widely differnt scales.
The first two generations are completely degenerate to each
other due to our choice of discrete symmetry while the third
can also be made almost degenerate with the first two generations
through some reasonable fine tuning
of the model parameter $\xi$. The present model also admits large transition
magnetic
moment $\sim {10}^{-11}\mu_B$ consistent with the present
experimental limits due to the charged Higgs exchange at the one loop level
through the oscillation of $\nu_{eL}\rightarrow \nu_{\mu R}$, which seems to be
a reasonable choice for the solution of the existing anticorrelation between
the Sun spot activity and the observed neutrino flux on earth but
evades the constraints from the SN 1987 A data and nucleosynthesis.
\vskip .5in
A.G. acknowledges financial support from  the UGC, Govt. of India.
\newpage
\begin{center}
{\Large\bf References}
\end{center}
\begin{enumerate}

\item  D.Caldwell and R.N.Mohapatra , Phys. Rev. D48, (1993) 3259.

\item  K.Lande et.al. Proc. XXVth Int. Conf. on High Energy Physics,
Singapore, ed. K. K. Phua  and  Y.  Yamaguchi,  World  scientific
(Singapore, 1991); K.S.  Hirata  et.al.  Phys.Rev.  Lett.  66,
(1991) 9; P. Anselmann et. al. Phys. Lett. B285, (1992)  376;
A. I.
Abazov et.al. Phys. Rev. Lett. 67,  (1991)  3332;  V. Garvin ,  talk
at Int. Conf. on HEP , Dallas 1992.

\item  E. M. Beier et.al.  Phys. Lett.  B280, (1992)  149 ; D. Casper
 et.al. Phys. Rev. Lett. 66, (1991) 2561.

\item  A. N. Taylor and M. Rowan Robinson, Nature  359, (1992)  396;
M. Davis , F. J. Summers and D. Schlegel, Nature 359, (1992) 393.

\item GALLEX Collaboration, P.Anselmann et.al. Phys. Lett. B342,
(1995)  440.

\item  J.N.Bachall, P.I.Krastev and E.Lisi, Phys. Lett. B348,
(1995) 121,
  E. Kh. Akhmedov, A. Lanza and S. T. Petcov,
Phys. Lett. B348, (1995)  124.

\item B.Achkar et.al. Nucl. Phys. B434 , (1995) 503.

\item D.R.O.Morrision, CERN-PPE/95-47, (1995).


\item  A.S.Joshipura, Z.Phys. C64, (1994) 31.

\item P.Bamert and C.P.Burgess, Phys. Lett. B329, (1994) 289.

\item K. S. Babu, IV Mexican school of Particles and Fields, ed.
 by J. Lucio M and A. Zepeda (World  Scientific,  Singapore,
1992), p 104;
  P. Langacker, Summer school in HEP and Cosmology, Trieste
 1992 ed. E. Gava et.al. (World  Scientific, Singapore) p 487;
 M. Fukugita and T. Yanagida, in "Physics and Astrophysics of
Neutrinos" ed. by M. Fukugita and A. Suzuki, (Springer Verlag, 1995).

\item M. Gell Mann,  P. Ramond and  R. Slansky in "Supergravity", ed.
 by  D. Z. Freedman, (North Holland, 1979), T. Yanagida, in
Proceedings of "Workshop on Unified theory and Baryon Number in
the Universe"  ed. by  O. Sawada and A. Sugamoto,  (KEK, 1979),
R.N. Mohapatra and G. Senjanovic, Phys.  Rev.  Lett. 44, (1980) 912
,  Phys. Rev. D23, (1981) 165.

\item L. B. Okun,  M. B. Voloshin and M. I. Vysotsky,  Sov.  J.  Nucl.
 Phys. 44, (1986) 440, M. B. Voloshin and M. I. Vysotsky,  Inst.
 for Theoretical and Experimental Physics (Moscow) Report No.
 86-1, (1986), L. B. Okun,  M. B. Voloshin  and  M. I. Vysotsky,
 Inst. for  Theoretical  and  Experimental Physics
(Moscow) Report No. 86-82, (1986).

\item C. L. Cowan Jr. and F. Reines, Phys. Rev. Lett.  107, (1957) 528;
J. Kim, V. Mathur and S. Okubo, Phys. Rev. D9, (1974) 3050 ;  A.  V.
Kyuldjiev, Nucl. Phys. B243 , (1984) 387.

\item W. J. Marciano et.al.  Particle Data Group , Phys. Lett.
B239, (1990) 1, P. Vogel and J. Engel, Phys. Rev. D39, (1989) 3378.

\item M. Fukugita and S. Yazaki, Phys. Rev. D36, (1987) 3817 ,
S. I. Blinnikov , Moscow preprint, ITEP-88-19 (1988); G. G. Raffelt,
Phys. Rev. Lett. 64, (1990) 2856.

\item K. S. Babu and R. N. Mohapatra, Phys. Rev. Lett. 64,
(1990) 1705, G. Ecker, W. Grimus and H. Neufeld, Phys. Lett. B232,
(1989) 217.

\item D.Chang, W.Y.Keung and G.Senjanovic, Phys.Rev. D45, (1992)
 31.

\item Y. Aharonov, I. Goldman, G. Alexander and S. Nussinov,
Phys.  Rev.  Lett.  60,  (1987)  1789, J. Lattimer and J.
Copperstein, Phys. Rev.  Lett. 61,
 (1988) 25, R. Barbieri and R. N. Mohapatra, ibid, 61 (1988)
 27;
D. Notzold Phys. Rev. D38, (1990) 1658.

\item Y. Chikashige, R. N. Mohapatra and R. D. Pecci, Phys. Rev.
Lett. 45, (1980) 1926, Phys. Lett. B98, (1981) 265.

\item E. Ma, Phys. Rev. D43, (1991) 587, (1991)  2761;
N. G. Deshpande, M. Gupta and P. B. Pal, Phys. Rev. D45,  953 (1992),
N. G. Deshpande and  X. G. He, OITS preprint-529/94.

\item G.Gelmini and E. Roulet, CERN-TH- 7541/94.

\item G.Gelmini and T.Yanagida, Phys. Lett. B294, (1992) 53.

\item V.Berezinsky and J.W.F.Valle, Phys. Letts. B318, (1993) 360,
J. T. Peltoniemi and J. W. F. Valle, Phys. Letts. B304, (1993) 147,
J. McDonald Phys. Letts. B323, (1994) 339.

\item Andre Sopczak, CERN-PPE/ 95-46, hep-ph/9504300.

\item C. S. Lim and W. Marciano, Phys. Rev. D37, (1988) 1368, E.
Kh. Akhmedov, Phys. Lett. B213, (1988) 64.

\end{enumerate}
\end{document}